\documentclass[prd, twocolumn,showpacs,floats,floatfix,nofootinbib]{revtex4}

   \usepackage{graphicx}
   \usepackage{epstopdf}
\usepackage{color}
\usepackage{natbib}
\usepackage{dcolumn}
\usepackage{amssymb}
\usepackage{amsmath}
\usepackage{bm}
\usepackage{amsfonts}
\usepackage{txfonts}
\usepackage{bbm}
\usepackage{bbold}

\usepackage[pdftex,left=1in,right=1in,top=1in,bottom=1in]{geometry}

\newcommand{\ex}{\mathrm{e}}
\newcommand{\dd}{\mathrm{d}}

\def\spose#1{\hbox to 0pt{#1\hss}}

\def\lta{\mathrel{\spose{\lower 3pt\hbox{$\mathchar"218$}}
     \raise 2.0pt\hbox{$\mathchar"13C$}}}
\def\gta{\mathrel{\spose{\lower 3pt\hbox{$\mathchar"218$}}
     \raise 2.0pt\hbox{$\mathchar"13E$}}}

\def\setR{\mathbb{R}}
\def\setC{\mathbb{C}}

\def\setZ{\mathbb{Z}}

\newcommand{\ie}{\textsl{i.e.~}}

\newcommand{\Lag}{\mathcal{L}}

\def\beq{\begin{equation}}
\def\eeq{\end{equation}}
\def\bea{\begin{eqnarray}}
\def\eea{\end{eqnarray}}
\def\eqref{\ref}

\def\cL{c_{_\mathrm{L}}}

\begin{document}
\title{Can type II Semi-local cosmic strings form?}

\author{Betti Hartmann}
\email{b.hartmann@jacobs-university.de}
\affiliation{School of Engineering and Science,
Jacobs University Bremen, 28759 Bremen, Germany}

\author{Patrick Peter}
\email{peter@iap.fr}
\affiliation{${\cal G}\setR\varepsilon\setC{\cal O}$ -- Institut
d'Astrophysique de Paris, UMR7095 CNRS, Universit\'e Pierre \& Marie Curie,
98 bis boulevard Arago, 75014 Paris, France}

\date{\today}

\begin{abstract}
We present the simplest possible model for a semi-local string
defect in which a U(1) gauged subgroup of an otherwise global
SU(2) is broken to produce local cosmic strings endowed with
current-carrying properties. Restricting attention to type II vortices
for which the non current-carrying state is unstable, we show
that a condensate must form microscopically and macroscopically
evolve towards a chiral configuration. It has been suggested that
such configurations could potentially exist in a stable state, thereby
inducing large cosmological consequences based on equilibrium angular
momentum supported loop configurations (vortons). Here we show
that the current itself induces a macroscopic (longitudinal) instability:
we conclude that type II semi-local cosmic strings cannot form in
a cosmological context.
\end{abstract}


\maketitle
	
\section{Introduction}

Cosmic strings have long lost their status of plausible competitors to
the inflation paradigm \cite{PPJPU}. However, from the point of view
of particle 
physics and high energy models thereof, the opposite should be true:
even though it is not immediately obvious to build consistent models
of inflation based on the most natural extensions of the standard model
such as supersymmetric Grand Unified Theories (GUT) or strings,
those naturally predict vortex-like objects, i.e. linear topological defects
\cite{JRS03} (see however Ref. \cite{Clesse11}).
Thus, constraints provided by cosmic string network simulations are
very much still of current interest, would it be only to understand why
and how one can construct an inflation model without strings.

Assuming strings to form however is not yet the end of the story. In
practice, most research has been made under the assumption that
the vortices were not endowed with any particular structure, and hence
that the spacelike two-dimensional worldsheet they described was
well modeled by a Nambu-Goto Lorentz invariant action, \ie the area
spanned by the worldsheet.

That such a model attracted attention makes full sense since it
turns out that any more complicated model would be essentially
intractable by means of the currently available technology. Besides,
it was also shown that any Lorentz symmetry-breaking current
on the vortices could lead to centrifugally-supported equilibrium
states, dubbed vortons \cite{DavisShellard1,DavisShellard2},
whose existence merely rules out the 
string scenario altogether \cite{vortons6},
provided they are sufficiently long-lived.

Structureless Nambu-Goto strings, on the other hand, are very difficult to
produce in almost any reasonable high energy theory. Indeed,
and unless one assumes a special sector put by hand to generate
the strings themselves, which comes very short of the original
idea to describe the high energy phenomena in a unified and
consistent way, the string-forming Higgs field present in most
GUT model must couple to scalars, fermions or gauge fields in
such a way as to produce currents. Even the cosmic strings
present in the superstring framework do not escape this conclusion,
as they must couple to moduli, at least the volume of the
compact extra dimensions. Thus, one expects cosmic strings to
be of the current-carrying kind, as originally introduced by Witten
in 1985 \cite{Witten85}.

Many models have since been discussed and investigated by
numerous authors, with the general conclusion that the equation
of state of the strings is highly non trivial, with specific properties
such as the existence of a maximal spacelike current, a phase
frequency threshold for timelike current above which there is
no bound state anymore, and the possibility, in all known models,
to build a lightlike current which ought to be absolutely stable,
thus enhancing the vorton excess problem
\cite{neutral,enon0,BettiBrandon08}. Solutions have been proposed,
most of them based on the
instabilities of current-carrying loop configurations that would
dissipate most of the large loops before they have time to
evolve into cosmologically dangerous vortons. The present
work, although not directly concerned with this problem, suggests
yet another possibility, namely that the current could form directly
in a configuration that would be unstable with respect to
longitudinal (soundlike along the string) modes.

Our model can be seen as the next-to-simple one after the
neutral Witten bosonic model, consisting of a global U(1)
condensate in a local U(1) vortex.
Here we still assume the
vortex to be produced by a gauged U(1) symmetry breaking,
but instead of adding extra symmetries, we embed this local
U(1) into an otherwise global SU(2). Non-current carrying
strings in this model have been investigated 
in \cite{Vachaspati:1991dz,Hindmarsh92,Hindmarsh:1992yy,Achucarro:1999it}, while
the current-carrying case has been discussed in \cite{ForgacsVolkov06,Forgacs2006}.
This is merely the limit
of the usual would-be semi-local strings found in the standard
electroweak model; except that the measured parameters of
this model preclude their actual stability.
In fact, the stability of non-current carrying semi-local strings does not 
follow from the topology of the vacuum manifold (as it does for
the U(1) case), 
but from dynamical arguments.

The ratio between the gauge and Higgs boson masses
governs the stability of semi-local strings: for Higgs boson mass larger (smaller) than 
the gauge boson mass semi-local strings are unstable (stable) and in the BPS limit a
degenerate one-parameter family of stable solutions exists \cite{Hindmarsh92}.
The parameter corresponds roughly to the width of the strings and as such semi-local strings of arbitrary width
have the same energy in the BPS limit. 
Whenever this zero mode gets excited it leads to the growth of the string core 
\cite{Leese:1992hi}. As such these non-current carrying semi-local strings 
have been studied in the context of cosmological applications
regarding the formation and evolution of string networks \cite{Achucarro:1992hs,Achucarro:1997cx,Achucarro:1998ux,
Achucarro:2007sp} as well as implications for the CMB \cite{Urrestilla:2007sf}. The stability
of the current-carrying counterparts has been discussed in \cite{GaraudVolkov08} using
linear perturbation theory; there it was also found that these
embedded type II vortices have a single unstable mode, and so it has
been suggested that the current-carrying ones, being less energetic, could be stable.
We show that this is not the case because some other instability develops.

In a sense, the category of this model is more natural
than the Witten-kind of models because one expects a large
GUT group to be partially broken to yield the low energy particle
physics currently tested at the LHC, so the strings, if present, once
formed, are expected to be embedded in a larger structure. It
is obviously mostly a parameter dependent question to know
whether the strings here described will form rather than the Witten
kind of strings. Finally, such a model permits to embed a
cosmic string in a non abelian framework in a tractable way,
contrary to what happens in the case of a pure non abelian
current-carrying situation \cite{prdFdMMLJMPP,NonAbCarter11}.

As already mentioned above, if the ratio between the Higgs and
gauge boson masses is large, the corresponding type II vortices are
unstable. In Ref.~\cite{ForgacsVolkov06} and \cite{Forgacs2006}, it was
shown that a current could build along such vortices, and that the
resulting current-carrying state was less energetic than the structureless
one. A stability analysis \cite{GaraudVolkov08} then showed that even
though long wavelength perturbations tend to grow exponentially, there
was a limit below which the current-carrying string state could be stable;
this could imply important cosmological consequences whenever small
loops form. The purpose of the current article is to close this window of
stability by performing a global analysis showing the current-carrying
configurations will also develop a short wavelength instability, the so-called
longitudinal instability introduced by Carter \cite{stabvorton,BCPP95,models}.

The paper is organized as follows: in the following section
\ref{sec:model}, we set up the actual model and discuss the
stringlike solutions that can be expected. We then move on,
in Sec.~\ref{sec:current} to evaluating the currents that could
condense in a string core, summarizing a stability analysis first
discussed in \cite{Hindmarsh92}. These
currents are examined thoroughly in Sec.~\ref{sec:light} and it
is shown that the lightlike current limit is defined as the endpoint
of the state parameter space in this case,
with the phase frequency threshold being at the null point. Finally,
Sec.~\ref{sec:eos} shows that the corresponding equation of
state leads to the longitudinal loop instabilities:
right after a condensate has formed,
it should evolve towards the chiral limit \cite{CP99}, thereby  destroying
many would-be vortons \cite{vortons7} through emission of high energy particles
\cite{MP00,3D}.
We conclude that type II vortices cannot form at all in such models.

\section{Partly gauged SU(2) string model} \label{sec:model}

The simplest  embedded current-carrying string model is
provided by the partly ungauged version of the electroweak
theory in which the SU(2) coupling constant is made to vanish,
while the equivalent to electromagnetism U(1) remains gauged.
In practice, this amounts to starting with the following Lagrangian
\begin{equation}
\Lag = -g^{\mu\nu} (D_\mu\bm{\Phi})^\dagger \cdot D_\nu\bm{\Phi}
-\frac14 F_{\mu\nu} F^{\mu\nu} - V(\bm{\Phi}),
\label{LagIni}
\end{equation}
where the U(1) covariant derivative acting on the SU(2) Higgs doublet
$\bm{\Phi}$ is $D_\mu\bm{\Phi} \equiv \left( \partial_\mu - i e A_\mu
\right) \bm{\Phi}$, $F_{\mu\nu} \equiv \partial_{\mu}A_{\nu} -
\partial_{\nu}A_{\mu}$
is the Faraday tensor of the U(1) gauge field, and finally the
scalar field potential $V$ is taken to be of the symmetry-breaking
kind
\begin{equation}
V(\bm{\Phi}) = \frac{\lambda}{2} \left( \bm{\Phi}^\dagger\cdot\bm{\Phi} 
-\eta^2 \right)^2,
\label{Vphi}
\end{equation}
so the self coupling $\lambda$ combines with the vacuum expectation
value (vev) $\eta$ of $\bm{\Phi}$ to provide the scalar field excitation
mass as $m_\phi = \sqrt{2\lambda} \eta$. The vector field also
acquires a mass $m_A = \sqrt{2} e\eta$, and the mass ratio is thus
defined as $2 \beta\equiv m_\phi^2 / m_A = \lambda/e^2$. (Note our
definition of $\beta$ differs by a factor of 2 with that of
Ref.~\cite{ForgacsVolkov06}.)

The lowest energy configuration, having $\bm{\Phi}^\dagger \cdot
\bm{\Phi} = \eta^2$ admits vortex defects of the local U(1) kind:
fixing the SU(2) gauge in which 
\begin{equation}
\bm{\Phi}_0 = \begin{pmatrix} \Phi_0 \\ 0 
\end{pmatrix},
\label{iniconf}
\end{equation}
there remains a local U(1) gauge to be fixed through the phase
of $\Phi_0$; if it takes the form of a non vanishing winding, \ie if
$\Phi_0 \propto \ex^{in\theta}$ with index $n\in\setZ\not= 0$ and
$\theta$ a local coordinate angle, then $\Phi_0\to 0$ defines a
string around which the phase winds. One can then locally set the
string to be aligned along a $z-$axis around which one defines
the cylindrical coordinates $r$ and $\theta$, and the non vanishing
component of the Higgs field becomes $\Phi_0 = \varphi(r)
\ex^{in\theta}$, where $\lim_{r\to\infty} \varphi(r)=\eta$ and
$\varphi(0)=0$. 

The question then arises as to the actual stability of the above
configuration. An analysis similar to that in \cite{Witten85} is carried
out below showing that one does indeed expect a current of
the kind we discussed in the following sections.

{}From the Lagrangian (\ref{LagIni}), one obtains the general
equations of motion for the gauge field $A_\mu$ as
\begin{equation}
\frac{1}{\sqrt{-g}}\partial_\nu \left(\sqrt{-g}F^{\nu\mu}\right) =
2 e^2 \bm{\Phi}^\dagger\cdot \bm{\Phi} A^\mu + i e \bm{\Phi}
\stackrel{\leftrightarrow}{\partial}\!{}^\mu \bm{\Phi},
\label{EOMA}
\end{equation}
and for the Higgs scalar
\begin{equation}
\frac{1}{\sqrt{-g}} \partial_\mu \left(\sqrt{-g}g^{\mu\nu}
D_\nu \bm{\Phi}\right) = i e A^\nu
D_\nu\bm{\Phi} +\bm{\Phi}
\frac{\dd V(\bm{\Phi})}{\dd (\bm{\Phi}^\dagger\cdot\bm{\Phi} )},
\label{EOMbmPhi}
\end{equation}
with the hermitian conjugate equation applying for
$\bm{\Phi}^\dagger$. These give, for the background
configuration (\ref{iniconf}) with the potential (\ref{Vphi}),
\begin{equation}
\frac{\dd^2\varphi}{\dd r^2}+\frac{1}{r}\frac{\dd \varphi}{\dd r}
= \left[ \frac{Q^2}{r^2} + 
\lambda \left( \varphi^2-\eta^2\right)
\right] \varphi,
\label{EOMvarphi}
\end{equation}
and
\begin{equation}
\frac{\dd^2 Q}{\dd r^2}-\frac{1}{r}\frac{\dd Q}{\dd r}
= 2e^2 \varphi^2 Q,
\label{EOMQ}
\end{equation}
after setting $Q=n-eA_\theta$ to account for the winding number. We
now assume -- see the following sections -- that we have (numerical)
solutions for the functions $\varphi (r)$ and $Q(r)$.

Because the Higgs doublet is coupled with itself, and even though
finite energy solutions of Eqs. (\ref{EOMvarphi}) and (\ref{EOMQ})
exist, one needs verify that these are stable. Following Witten \cite{Witten85},
we set an arbitrary perturbation $\bm{\Phi} = \bm{\Phi}_0 +
\delta\bm{\Phi}$ with
\begin{equation}
\delta\bm{\Phi} = \begin{pmatrix} 0 \\ \sigma \ex^{i\omega t}
\end{pmatrix},
\label{deltaPhi}
\end{equation}
where $\sigma=\sigma(r)$ depends on the radial coordinate only. Plugging
Eq. (\ref{deltaPhi}) into (\ref{EOMbmPhi}) and keeping only first order
terms, one gets the Schr\"{o}dinger-like equation
\begin{equation}
-\Delta_2 \sigma + \mathcal{V}(r) \sigma = \omega^2 \sigma,
\label{SchrSigma}
\end{equation}
where $\Delta_2 = \partial_x^2 + \partial_y^2=\partial^2_r + r^{-1} \partial_r
+r^{-2} \partial_\theta^2$ is the two-dimensional laplacian and the
effective potential $\mathcal{V}$ reads
\begin{equation}
\mathcal{V}(r) = \frac{\left[n-Q(r)\right]^2}{r^2}+ \lambda \left[
\varphi^2(r) -\eta^2\right].
\label{calV}
\end{equation}

\begin{figure}[ht]
\centering
\includegraphics[width=8cm,clip]{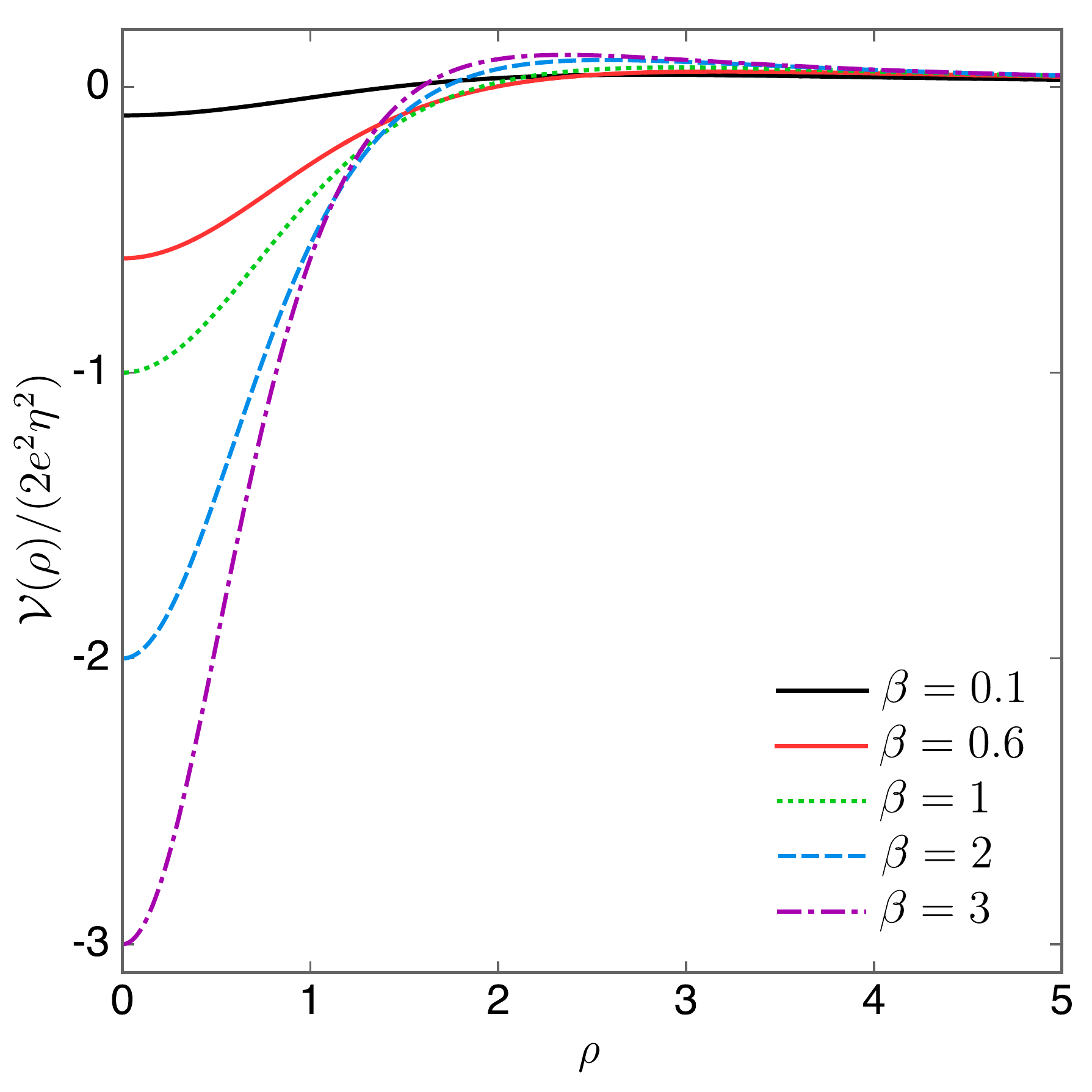}
\caption{The potential (\ref{calV}), rescaled so as
to be dimensionless, appearing in the Schr\"odinger equation
(\ref{SchrSigma}) for various values of the parameter $\beta=
\lambda/(2e^2)$ as a function of the dimensionless distance
to the string core $\rho=\sqrt{2} e\eta r$ 
(see Sec. \ref{sec:num} for details on the numerics).}
\label{fig:Pot}
\end{figure}

This potential is shown on Fig.~\ref{fig:Pot} for different values of the
parameter $\beta\equiv\lambda/(2e^2)$. One expects from the figure
that there could be bound states provided $\beta$ is large enough.

Since $\lim_{r\to 0}\mathcal{V}(r) = -\lambda\eta^2$ is negative
and $\mathcal{V} \sim n^2/r^2$ asymptotically, the potential satisfies
the usual quantum mechanical
conditions for having a bound state: a range of
values for the parameter $\beta$ can be
found for which there exist solutions of Eq. (\ref{SchrSigma})
with $\omega^2 <0$, and hence an instability of the background
solution (\ref{iniconf}) should develop. With the non linear terms
taken fully into account, the instability translates into a condensate
that can carry a current. Comparison with Ref. \cite{Hindmarsh92}
shows that for $\beta>\frac12$, \ie $\lambda>e^2$, one expects
a condensate to form: according to the usual classification, this
means that type I vortices are absolutely stable (no condensate)
while type II vortices spontaneously form a current-carrying
state. Note also that since type II vortices are energetically
favored to occur with unit winding number, we shall for now
on restrict attention to the case $n=1$. The question now
is whether or not these current-carrying solutions can lead to
the stable enough configurations (for cosmological purposes)
discussed in \cite{GaraudVolkov08}.

It should be remarked at this stage that the mere existence of
an instability does not guarantee that it has an endpoint which
one then identifies with the current-carrying state. The numerical
solutions obtained below show that it does, and because the
field equations stem from minimizing the energy per unit
length to be minimized, they provide more stable configurations
satisfying the boundary conditions. As we shall see, these
solutions will turn out to initiate another instability.

\section{The current-carrier condensate} \label{sec:current}

For now on, we follow \cite{GaraudVolkov08} and
assume a condensate did form and we write the
Higgs doublet as
\begin{equation}
\bm{\Phi} =\left[ \begin{matrix} \varphi(r) \ex^{in\theta + i\psi(z,t)} \\
\sigma (r)\ex^{im\theta + i \xi(z,t)}
\end{matrix} \right],
\label{deltaPhi}
\end{equation}
where $n\in \setZ$ is the winding number of the string, $m\in \setZ$
leaves the possibility for the perturbation to wind as well, and the phases
$\psi$ and $\xi$ only depend on the internal string coordinates. This
field can then source $A_\theta$, $A_z$ and $A_t$, all being functions
of the radius $r$ only in order for the worldsheet to be localized. Note that
the form (\ref{deltaPhi}) assumes no modes are present in the
transverse direction, \ie the phases $\psi$ and $\xi$ do not depend
on $r$, so we do consider neither ingoing nor outgoing waves: the
field configuration we are investigating is at equilibrium, hence may
only have excitations along the worldsheet. We shall also occasionally
use a latin index to denote worldsheet coordinates $\{z,t\}$ collectively.

\subsection{State parameters}

With the ansatz (\ref{deltaPhi}), the field equations now read
\begin{equation}
A''_a+\frac{1}{r}A'_a +2 e \left[ (\partial_a \psi -eA_a ) \varphi^2
+ (\partial_a \xi -eA_a ) \sigma^2 \right] = 0,
\label{EOMAa}
\end{equation}
for the internal gauge fields,
\begin{equation}
Q''-\frac{1}{r}Q'=2 e^2 \left[ Q \varphi^2
+ (Q+m-n) \sigma^2 \right] ,
\label{EOMQtot}
\end{equation}
with the same convention as before that $Q=n-eA_\theta$,
\begin{equation}
\varphi''+\frac{1}{r}\varphi' =  \left[ P_\psi^2 +
\frac{Q^2}{r^2} +\lambda \left( \varphi^2+\sigma^2-\eta^2\right) 
\right]\varphi,
\label{EOvarphitot}
\end{equation}
with $P_\psi^2 = (\partial_z\psi-eA_z)^2 - (\partial_t\psi-eA_t)^2$,
\begin{equation}
\sigma''+\frac{1}{r}\sigma' =  \left[ P_\xi^2 +
\frac{(Q+m-n)^2}{r^2} +\lambda \left( \varphi^2+\sigma^2-\eta^2\right) 
\right]\sigma,
\label{EOMsigmatot}
\end{equation}
where $P_\xi^2$ is defined in a similar fashion as $P_\psi$, namely
$P_\xi^2 = (\partial_z\xi-eA_z)^2 - (\partial_t\xi-eA_t)^2$.
Finally, the phases represent massless modes propagating along the
string, as is clear from their equations of motion
\begin{equation}
(\partial^2_t - \partial^2_z)\psi = \gamma^{ab}\partial_a \partial_b
\psi =0 = \gamma^{ab}\partial_a \partial_b \xi.
\label{EOMphases}
\end{equation}
In Eqs. (\ref{EOMAa}) to (\ref{EOMphases}), we have set a prime
to denote a derivative with respect to the radial distance $r$.

One now needs to look at the boundary conditions to restrict
attention to the physically meaningful cases. In particular, noting
that $\lim_{r\to 0}Q(r)=n$ and assuming $P_\xi^2$ to be regular
at the string core location, Eq. (\ref{EOMsigmatot}) implies the
following: setting $\sigma \simeq \sigma_0 + \sigma_0' r +\frac12
\sigma_0'' r^2 +\cdots$, the expansion
$$
\sigma_0'' \left( 2-\frac{m^2}{2}\right) + \frac{\sigma_0'}{r}
\left( 1-m^2 \right) + \frac{m^2 \sigma_0}{r^2} P_\xi^2(0)\sigma_0
+ \mathcal{O}\left( r\right)=0
$$
should hold. In order for the $r^{-2}$ term to be regular, one must
impose either $m=0$ or demand that $\sigma_0=0$. In the latter case,
assuming $m\not=0$, one finds that $m^2=1$ and $m^2=4$
simultaneously, which is self-contradictory. Hence, we must set
$m=0$ and $\lim_{r\to 0} \sigma'(r)=0$.
Moreover, asymptotically, \ie when $Q\to 0$,
$\sigma\to 0$ and $\varphi\to \eta$, Eq. (\ref{EOMAa}) becomes
\begin{equation}
A_a''+\frac{1}{r} A_a' +2e\eta^2 (\partial_a\psi - eA_a)=0,
\label{12prime}
\end{equation}
the solution of which can only be made to vanish -- \ie we
demand $\lim_{r\to\infty}A_a(r)=0$ in order for the total
energy of the configuration to be finite -- only provided
$\partial_a\psi=0$. As $\psi$ must now be a constant, it
can, without lacking generality, be set to zero by means
of a global SU(2) gauge transformation (which can also
remove any constant part that would be present in $\xi$
as well). The general solution of (\ref{EOMphases}) then
reads
\begin{equation}
\xi = \xi_-(z-t) + \xi_+(z+t) + kz-\omega t,
\label{omegak}
\end{equation}
where $\xi_\pm$ represent the left and right massless modes
moving along the string and the last term represents a coherent
mode, that can, in the usual case, be built as a superposition of left
and right movers. If a string segment is considered, the
left and right moving modes are responsible for the leaking out
of the current; again, following \cite{GaraudVolkov08}, we shall in
what follows consider a $z-$independent string (approximating a
closed loop when setting periodic boundary conditions), assuming
it can somehow be formed in the first place and thus neglect
these modes; we shall accordingly
set $\xi_\pm\to 0$ in what follows.

Because of Eq.~(\ref{omegak}), the last term of Eq.~(\ref{12prime})
is a constant. This implies that the two functions
$P_a \equiv eA_a-\partial_a\xi$ satisfy the same
linear equation and hence are merely proportional to one
another for all values of $r$. One then has $P_z \propto P_t$,
the proportionality constant being found by taking the asymptotic
limit of this relation for which we want the gauge field $A_a$ to
vanish. This yields $P_z=-k P_t /\omega$, and thus 
$A_z=-k A_t /\omega$. We are now in a position to define the relevant
degree of freedom as
$$
A_z^2 - A_t^2 = \left( \frac{k^2}{\omega^2}-1\right) A_t^2
=\left( 1-\frac{\omega^2}{k^2}\right) A_z^2 \equiv
w P^2,
$$
with $w$ the state parameter, and the function $P$ is dimensionless.
The fields $A_z$ and $A_t$ are then related to $P$ through
$$
A_t = \omega P \sqrt{\frac{w}{k^2-\omega^2}}, \ \ \ \ \hbox{and} 
\ \ \ \ A_z = -k  P \sqrt{\frac{w}{k^2-\omega^2}};
$$
note that $w$ has dimensions of a squared mass.
In view of this, one needs to complement the system with yet
another independent -- and dimensionless --
parameter $b$, representing the bias
between the gauge fields and the phase gradient, through
$$
k^2 - \omega^2 = w b^2.
$$
The sign of $w$ determines that of the phase gradient,
so the current is described by two positive parameters and a sign.
For $w>0$ (resp. $w<0$), the current is spacelike (resp. timelike),
and the equation of motion for $P$ is
\begin{equation}
P''+\frac{1}{r} P' = 2e^2 P \left( \varphi^2 + \sigma^2 \right) +2eb \sigma^2,
\label{EOMAs}
\end{equation}
where we assume $b>0$.

Having constructed the current-carrying configuration and taken
account of all the symmetries, we now turn to the range of parameters
that one should investigate to fully describe such strings.

\subsection{The lightlike current limit}
\label{sec:light}

The ordinary -- neutral \cite{neutralPP92} or charged \cite{enon0PP92} -- 
current-carrying cosmic string is known to have a maximum charge
density (timelike current) above which it is energetically favored for
the condensed particles to form ingoing and outgoing massive
radial modes. In the model here discussed, such a phase frequency
threshold is also acting, and as it turns out, it prevents the timelike
currents to form altogether.

With the degrees of freedom as obtained in the previous section, we
can rewrite Eq. (\ref{EOMsigmatot}) as
\begin{equation}
\sigma''+\frac{1}{r} \sigma' = \left[ P_\xi^2+ \frac{(Q-n)^2}{r^2}
+\lambda\left(\varphi^2+\sigma^2-\eta^2\right)\right]\sigma,
\label{sigsimpl}
\end{equation}
where now $P_\xi^2 = w\left( b+e P \right)^2$. In the asymptotic
regime, one is left with
\begin{equation}
\sigma''+\frac{1}{r}\sigma' \sim \left( w b^2 +\frac{n^2}{r^2}\right)
\sigma,
\label{SigmaBessel}
\end{equation}
as $\sigma$ decreases to vanishingly small values. The general
solution for this Bessel equation is
\begin{equation}
\sigma \sim a_\mathrm{I} I_n(b\sqrt{w} r) + a_\mathrm{K}
K_n(b\sqrt{w} r)
\label{Kn}
\end{equation}
for constant $a_\mathrm{I}$ and $a_\mathrm{K}$, with
$I_n$ and $K_n$ the modified Bessel functions of order $n$. For
$w>0$, the field is a condensate provided we set
$a_\mathrm{I}=0$.

The energy contained in this solution converges exponentially
fast far from the string core provided $w>0$: for $w<0$, instead
the general solution is a combination of oscillatory Bessel
functions. In the usual Witten current-carrying case
\cite{neutralPP92,BCPP95}, there is a similar transition for a given,
nonzero negative value $w_\mathrm{th}$ of $w$ that leads to a
logarithmic divergence in the equation of state in the limit $w\to
w_\mathrm{th}$. Here however, the threshold would be for a lightlike
current, with $w_\mathrm{th}=0$: the would-be divergence is
regularized by the $w$ prefactor that enters into the definition of
the energy per unit length and tension (see next Sec. \ref{sec:eos})
and the result is perfectly finite: there is no phase frequency
threshold\footnote{Rather, one could say that there is a frequency threshold
as in the usual case, but the asymptotic mass of the current-carrier vanishes
since it is akin to a Goldstone mode here, so the threshold does not imply
a divergent behavior of either the energy per unit length or the tension.}
in this case, the current can, from a spacelike configuration,
smoothly evolve towards an almost lightlike situation.

In fact, Eq. (\ref{Kn}) also gives the behavior of $\sigma$ with $w$
in the limit $w\to 0$. First, setting $w=0$ into Eq.~(\ref{SigmaBessel})
yields $\sigma \sim A r^{-n} + B r^{n}$, with $A$ and $B$ unknown
constants; a necessary condition for the condensate to be localized on
the vortex is that $B=0$. On the other hand, taking directly the solution
(\ref{Kn}) with $a_\mathrm{I}=0$ and expanding the Bessel function
$K_n$ in the neighborhood of $w\sim 0$ (we assume an analytic
continuation with $n\to n+\epsilon$ and take afterwards the limit
$\epsilon\to 0$ to handle the singularity), one obtains 
$$
K_n(b\sqrt{w} r) \sim 2^{-1-n} b^n r^n w^{n/2} \Gamma(-n)
+2^{-1+n} b^{-n} r^{-n} w^{-n/2} \Gamma(n),
$$
so that, providing $w^{n/2}$ converges to zero faster than
the pole in the $\Gamma$ function, one can identify
$$
a_\mathrm{K} = \frac{\left(b\sqrt{w}\right)^n}{2{n-1}(n-1)!} A,
$$
where $A$, although arbitrary at this stage, is independent
of $w$ as it comes from the solution for $w=0$. Therefore,
in the small (but finite) $w$ limit, we have that $\sigma \propto
w^{n/2} K_n(b\sqrt{w} r)$ whose asymptotic behavior gives
$\sigma\propto w^{n/2-1/4} \ex^{-b\sqrt{w} r} /\sqrt{r}$. It is
this behavior that implies the chiral current limit to be well
defined.

We now move on to evaluating the integrated quantities
leading to this equation of state.

\section{Integrated quantities} \label{sec:eos}

In order to describe the network of strings that will be generated
by the single strings here considered, one needs to integrate over
the transverse directions in order to be able to approximate each
defect by means of an actually zero thickness object. This means
we should derive the current and stress energy tensor associated
with the solutions obtained above.
As we will then show the string to be unstable with respect to
longitudinal perturbations, the worldsheet these integrated quantities
suppose will not actually last; assuming its presence is however
necessary for calculation purposes.

\subsection{Current}\label{sec:eosA}

Among the integrated quantities of interest, the current, defined as
\begin{equation}
J^\mu \equiv \frac{1}{2e} \frac{\delta\mathcal{L}}{\delta A_\mu},
\label{JmuDef}
\end{equation}
provides two independent ways to verify that the following
configurations obtained numerically are indeed solutions and
not mere artifacts. 
With the framework of model (\ref{LagIni}), this is
\begin{equation}
J^\mu = -\frac{i}{2} \left[ \bm{\Phi}^\dagger \cdot \left(\partial^\mu \bm{\Phi}
\right) -\left( \partial^\mu \bm{\Phi}^\dagger\right) \cdot \bm{\Phi}\right]
-e A^\mu \bm{\Phi}^\dagger\cdot \bm{\Phi},
\label{JmuPhi}
\end{equation}
which gives, using the explicit form (\ref{deltaPhi}) in terms of
the components of $\bm{\Phi}$
\begin{equation}
J_r = 0 \ \ \ \hbox{and} \ \ \ \ \ J_\theta = Q\varphi^2 + (Q-n) \sigma^2,
\label{JrJtheta}
\end{equation}
for the transverse components, and
\begin{equation}
J_a = \sigma^2 \left(\partial_a \xi-eA_a\right) -2e\varphi^2 A_a,
\label{Ja}
\end{equation}
with $a\in \{z,t\}$ for the longitudinal, worldsheet components.

Integration over the transverse degrees of freedom yield two
macroscopically defined quantities, namely the rotational current
flux around the string 
\begin{equation}
I_\theta \equiv \int\dd^2x^\perp J_\theta = 2\pi\int \left[ Q \varphi^2
+(Q-n)\sigma^2\right]\,r\,\dd r = \frac{2\pi n}{e^2},
\label{Ith}
\end{equation}
when the field equation (\ref{EOMQtot}) with $m=0$ is used, 
and the Lorentz-invariant current scalar $J$ along the worldsheet
defined through
\begin{equation}
J^2 \equiv \left( \int\dd^2x^\perp J_z \right)^2 - 
\left( \int\dd^2x^\perp J_t \right)^2 ,
\label{Jdef}
\end{equation}
which is readily evaluated in terms of the underlying field solution
previously derives as
\begin{equation}
J=2\pi\sqrt{w}\int \left[ eP\varphi^2 +\left( b+eP\right) \sigma^2 \right] \,r
\,\dd r,
\label{Jfields}
\end{equation}
because the difference of the integrals is itself a squared integral,
as expected for Lorentz symmetry reasons along the string. Making use
of the field equation (\ref{EOMAs}) then yields $J=0$, so this definition
cannot account for a conserved current along the worldsheet. This
stems from the fact that the current is now supported by both components
of the doublet, whereas in the usual Witten situation, there is only
one field that carries the current.

Although mostly useless for physical purposes, the current components
(\ref{Ith}) and (\ref{Jfields}) can be used as a measure of the validity of
the numerical calculation: once the fields are calculated, evaluating the
integrals should reproduce the analytic results above.

An alternative way to define the current is obtained by recalling that
it physically comes from the phase gradient along the string. In other
words, what really matters is the current-carrying {\em phase} instead
of the field itself, so that a suitable worldsheet covariant -- but not
SU(2) covariant -- definition is
\begin{equation}
\mathcal{J}_a = -\frac12 \eta_{ab}
\frac{\delta\mathcal{L}}{\delta\partial_b\xi},
\label{DefI}
\end{equation}
where $\eta_{ab}\equiv \mathrm{diag}\, (-1,1)$ is the internal Minkowski
metric in the string. Since the action only depends on the phase gradient
and not on the phase itself, this current is automatically conserved. With the
definition (\ref{DefI}), one can construct an integrated current $I$ which
is merely one part of that given in (\ref{Jfields}), namely one finds,
using the same integration procedure as in (\ref{Jdef}) (with the
replacements $J\to I$ and $J_a\to\mathcal{J}_a$)
\begin{equation}
I=2\pi\sqrt{w}\int \left( b+eP\right) \sigma^2 \,r \,\dd r.
\label{Ifields}
\end{equation}
The nonzero value of this quantity also explains the difference
between the spacelike and timelike eigenvalues of the stress 
energy tensor to which we now turn.

\subsection{Worldsheet stress-energy tensor}

From the Lagrangian (\ref{LagIni}), one also derives the
stress energy tensor
\begin{equation}
T_{\mu\nu} = -2\frac{\delta \mathcal{L}}{\delta g^{\mu\nu}}
+g_{\mu\nu} \mathcal{L},
\label{Tmunu}
\end{equation}
leading to the worldsheet components
\begin{eqnarray}
T_{tt} &=&2e^2 \varphi^2 A_t^2 + 2\sigma^2 \left(\partial_t\xi 
-e A_t \right)^2 + A_t'^2
-\mathcal{L}(\varphi,\sigma,Q,P),\cr && \label{Ttt}\\
T_{zz} &=& 2e^2 \varphi^2 A_z^2 + 2\sigma^2 \left(\partial_z\xi 
-e A_z \right)^2+A_z'^2
+\mathcal{L}(\varphi,\sigma,Q,P),\cr && \label{Tzz}\\
T_{zt} &=& 2e^2 \varphi^2 A_z A_t + 2\sigma^2 \left( \partial_z \xi
-e A_z \right) \left( \partial_t -e A_t \right) + A_z'A'_t,\cr && \label{Tzt}
\end{eqnarray}
where we have made use of the symmetries discussed in the
previous sections, and the Lorentz-invariant part stems from
the background Lagrangian
\begin{widetext}
\begin{equation}
\mathcal{L}(\varphi,\sigma,Q,P)=-\varphi'^2-\sigma'^2
-\frac{Q^2 \varphi^2}{r^2}-\frac{(Q-n)^2 \sigma^2}{r^2}
- w\left[ e^2\varphi^2 P^2 + (b+eP)^2 \sigma^2\right]
 -\frac12\left( w P'^2
+\frac{Q'^2}{e^2 r^2}\right)
-\frac{\lambda}{2}\left( \varphi^2 + \sigma^2 -\eta^2
\right)^2.
\label{LabFields}
\end{equation}
\end{widetext}
%\begin{eqnarray}
%\mathcal{L}(\varphi,\sigma,Q,A)&=&-\varphi'^2-\sigma'^2
%-\frac{Q^2 \varphi^2}{r^2}-\frac{(Q-m)^2 \sigma^2}{r^2} \cr
%& & - wb^2 \sigma^2 -\frac12\left( w A'^2
%+\frac{Q^2}{e^2 r^2}\right)\cr
%& & -\frac{\lambda}{2}\left( \varphi^2 + \sigma^2 -\eta^2
%\right)^2.
%\label{LabFields}
%\end{eqnarray}
We assume the other components, \ie in the transverse
direction, to vanish once integrated along the radial coordinates
for the on-shell solution \cite{PPcomment}. Following
\cite{LPXcoupled}, we write
%\begin{equation}
%T_{ab} = \left[ \begin{matrix} \mathcal{A} +\mathcal{B}
%+  & \cr 
%&\cr
%2\sigma k\omega + A_z' A_t'
% & -\mathcal{C}+\varphi^2 e^2 (A_z^2+A_t^2) + 
%\sigma^2 \left[ \left( k-eA_z\right)^2+\left(\omega+eA_t\right)^2 \right]
% +\frac12 (A_z'^2 
%+A_t'^2) \end{matrix}\right],
%\label{Tab}
%\end{equation}
\begin{equation}
T_{ab} = \begin{pmatrix} \mathcal{A}+\mathcal{B} & \mathcal{C} \cr
\mathcal{C} & -\mathcal{A}+\mathcal{B} \end{pmatrix},
\label{Tab}
\end{equation}
where $\mathcal{A}=-\mathcal{L}(\varphi,\sigma,Q,P;w\to0)$, \ie that part
of $\mathcal{L}$ of Eq. (\ref{LabFields}) without the variations
along the vortex, and 
\begin{eqnarray}
\mathcal{B} = & &\hskip-3mm\varphi^2 e^2 (A_z^2+A_t^2) + 
\sigma^2 \left[ \left( k-eA_z\right)^2
+\left(\omega+eA_t\right)^2 \right]\cr
& &\hskip-3mm+\frac12 (A_z'^2 +A_t'^2),
\label{calB}
\end{eqnarray}
and the non diagonal component reads
\begin{equation}
\mathcal{C} = 2\varphi^2 e^2 A_z A_t - 2\sigma^2 (k-e A_z) (\omega
+e A_t) + A_z' A_t' .
\label{calC}
\end{equation}
Diagonalization of $T_{ab}$ with respect to $\eta_{ab} = \mathrm{diag}
\, (-1,1)$ the two-dimensional Minkowski metric yields the eigenvalues
$E_\pm$.
Those are
\begin{eqnarray}
E_\pm &\equiv& \mathcal{A}\pm \sqrt{\mathcal{B}^2-\mathcal{C}^2}\cr
&=& \mathcal{A} \pm w \left[\frac12 P'^2+e^2 P^2 \varphi^2 + 
\left(b+e P\right)^2
\sigma^2 \right], 
\label{Epm}
\end{eqnarray}
from which one derives the energy per unit length $U$ and tension $T$
by integration over the transverse degrees of freedom, namely
\begin{equation}
U = 2\pi \int E_+(r) \, r\,\dd r \ \ \ \ \hbox{and} \ \ \ \ 
T = 2\pi \int E_- (r) \, r\,\dd r.
\label{UT}
\end{equation}
Note at this point that since the quantity appearing in the 
diagonalizing solution Eq. (\ref{Epm}) is a perfect square,
the integration and diagonalization procedures commute,
just as in the case of the current for which (\ref{Jfields})
could be straightforwardly derived, so the resulting macroscopic
quantities are really defined in an unambiguous way.

\begin{figure}[t]
\centering
\includegraphics[width=8cm,clip]{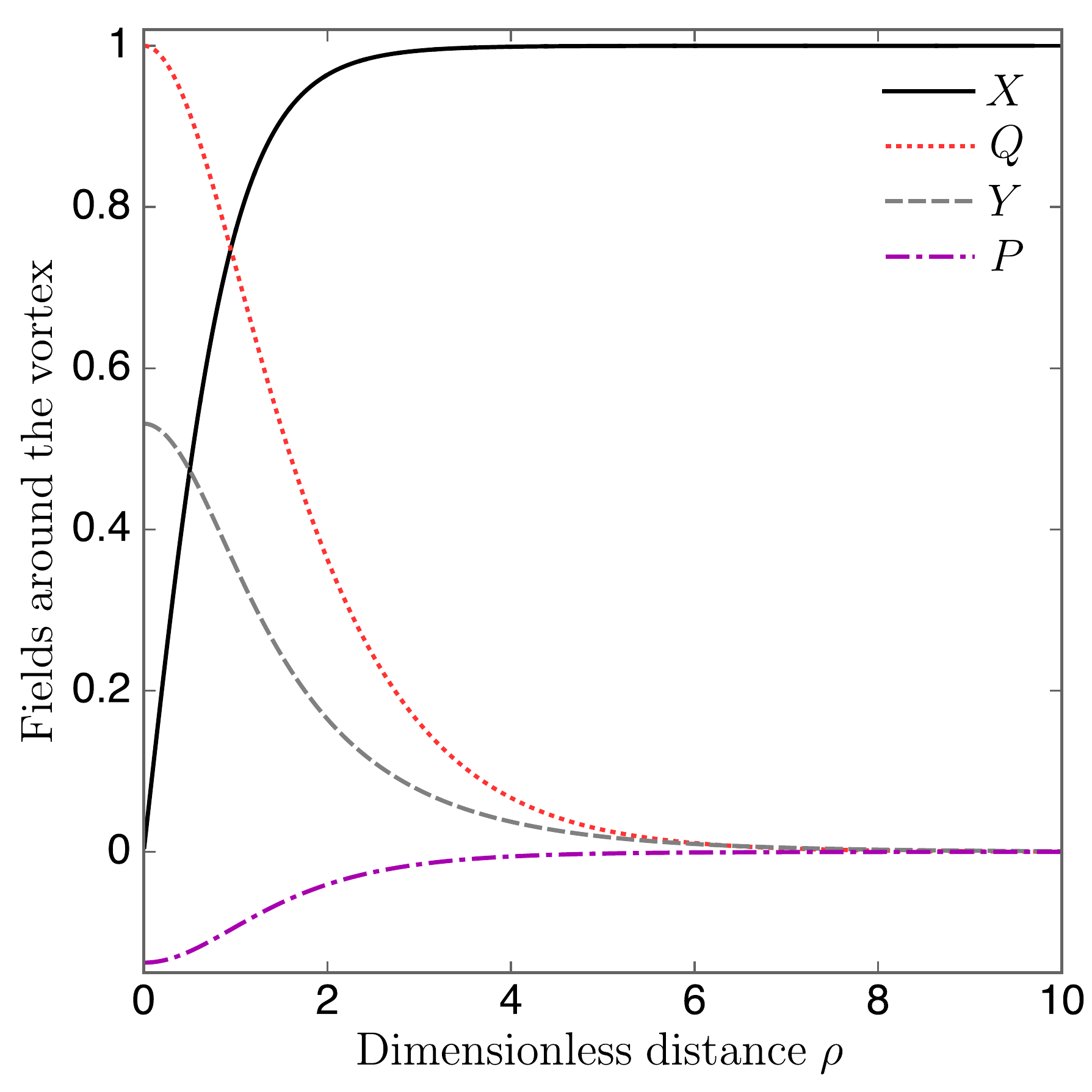}
\caption{Rescaled fields around the vortex: $X(\rho)$ -- full line -- 
and $Y(\rho)$ -- dashed --
are the Higgs field components in units of the Higgs VEV $\eta$, while
the vector field flux
$Q(\rho)$ -- dotted -- renders the vortex local and $P(\rho)$ -- dot-dashed --
condenses in such a
way as to support the current otherwise induced by condensation of
$Y$. This figure
is obtained for parameter values $\alpha=1$, $\beta=3$ and
$\tilde w=0.1 \beta/\alpha^2$.}
\label{fig:fields}
\end{figure}

In order to evaluate the actual behavior of the equation of state
relating the energy per unit length and the tension, and in particular
the stability of the resulting current-carrying string, we now discuss
the numerical solutions.

\subsection{Numerics}\label{sec:num}

Solving numerically the system of equations (\ref{EOMQtot}),
(\ref{EOvarphitot}), (\ref{EOMAs}) and (\ref{sigsimpl}), requires that
we cancel out the dimensions of the relevant quantities. Setting
$\rho =\sqrt{2} e\eta r$ the radius in units of the gauge vector mass, and
rescaling the fields and state parameter through
$$
\varphi = \eta X(\rho), \ \ \ \ \sigma = \eta Y(\rho) \ \ \ \ \hbox{and}
\ \ \ \ w = 2 \eta^2 \tilde w,
$$
we obtain the dimensionless equations of motion in the form
\begin{eqnarray}
\ddot X + \frac{1}{\rho}\dot X &=&\left[\tilde wP^2+\frac{Q^2}{\rho^2}
+\beta \left(X^2+Y^2-1\right) \right] X,\label{Xrho}\\
\ddot Q - \frac{1}{\rho}\dot Q &=& Q X^2+(Q-n) Y^2,\label{Qrho}\\
\ddot Y + \frac{1}{\rho}\dot Y &=&\left[\tilde w(\alpha+P)^2
+\frac{(Q-n)^2}{\rho^2}
+\beta \left(X^2+Y^2-1\right) \right] Y,\cr
& &\label{Yrho}\\
\ddot P + \frac{1}{\rho}\dot P &=& P\left( X^2+Y^2\right) +\alpha Y^2,
\label{Prho}
\end{eqnarray}
where a dot denotes differentiation with respect to the rescaled radius
$\rho$ and the constants are defined by $\alpha \equiv b/e$ and
$\beta\equiv\lambda/(2e^2)$. 

\begin{figure}[t]
\centering
\includegraphics[width=8cm,clip]{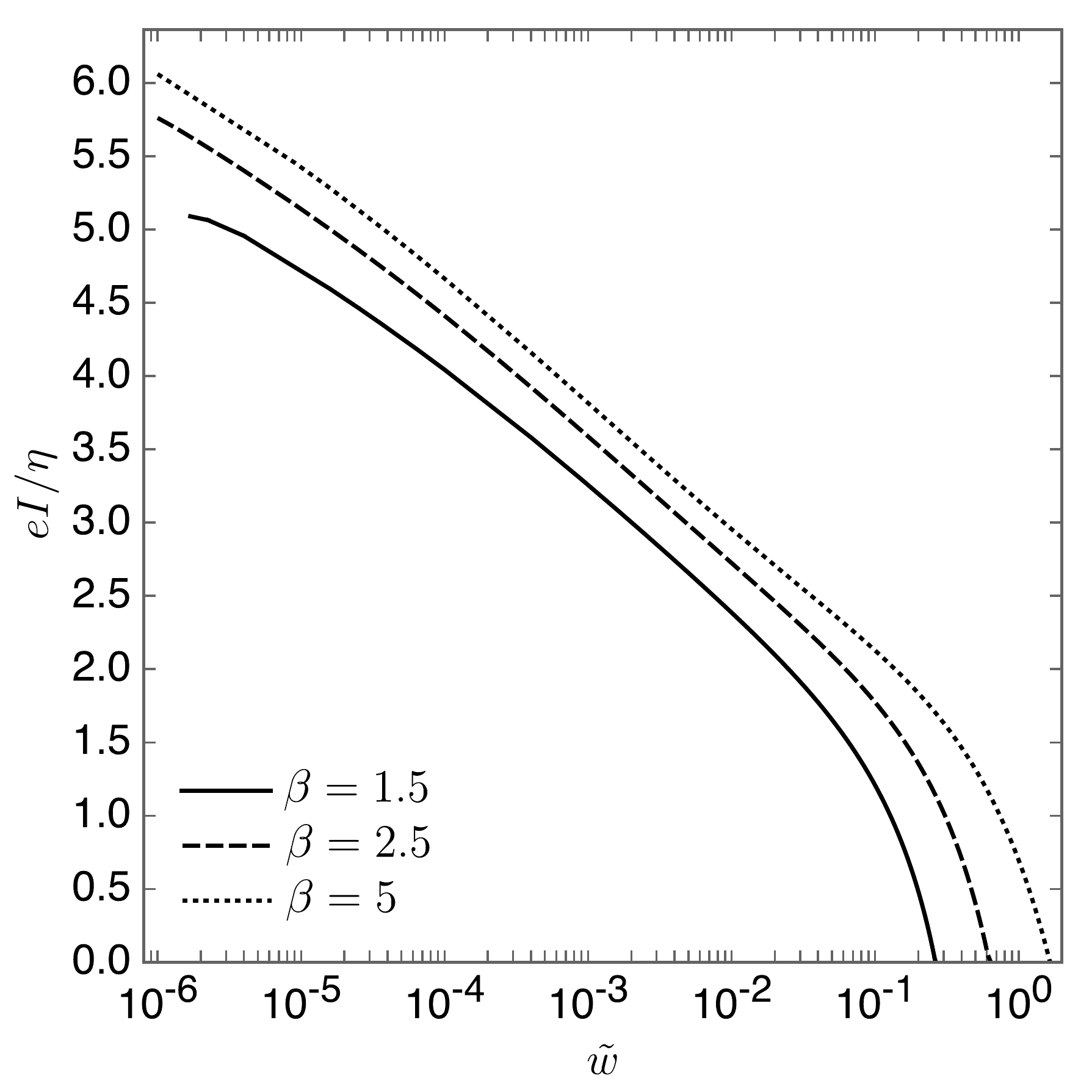}
\caption{Variation of the internal current $eI/\eta$
as a function of the rescaled state parameter
$\tilde w$ for $\alpha=1$ and various values of $\beta$ (same as on
Fig.~\ref{fig:UT}).}
\label{fig:IthI}
\end{figure}

A point worth discussing in relation with these equations concerns
the evolution of the condensate as the state parameter increases.
Expanding the field functions around the string core as
$X\propto \rho^m + \cdots$,
$Y\sim Y_0 + \frac12 \ddot X_0 \rho^2 +\cdots$, $Q\sim n
+ \frac12 \ddot Q_0 \rho^2 +\cdots$ and $P\sim P_0 + \frac12
\ddot P_0 \rho^2 +\cdots$, where we have taken into account
the regular boundary conditions, the zeroth order expansion
of Eqs.~(\ref{Xrho}) to (\ref{Prho}), one gets that
\begin{equation}
\ddot P_0 = \frac{Y_0^2}{2}\left(\alpha+P_0\right),
\label{ddP0}
\end{equation}
implying that $-\alpha\leq P_0\leq 0$: if $P_0> 0$, then (\ref{ddP0})
implies that $\ddot P_0>0$, and hence $P$ should be a growing and
positive function of $\rho$, which is inconsistent with the requirement
that $\lim_{\rho\to\infty} P= 0$ (we assume, following the figures, that
the functions are monotonic). If $P_0< -\alpha$, then $\ddot P_0 <0$,
the same argument applies with a negative and decreasing function.

Eq.~(\ref{Xrho}) tells us that $m=n$, as usual, while Eq.~(\ref{Qrho})
is trivially satisfied at the lowest order with the given expansion. However,
Eq.~(\ref{Yrho}) translates into
$$
\ddot Y_0 = \frac{Y_0}{2}\left[ \tilde w \left(\alpha+P_0\right)^2
+\beta\left( Y_0^2 - 1\right) \right],
$$
so that, demanding $\ddot Y_0 Y_0<0$ for the reasons just discussed
for $P$, one finds that
\begin{equation}
Y_0^2 \leq 1-\frac{\tilde w}{\beta}\left(\alpha+P_0\right)^2,
\label{Y0max}
\end{equation}
indicating that for large values of $\tilde w$, assuming $P_0$
to depend only mildly on $\tilde w$ (indeed, $P_0\to 0$ in this
limit), the available range for $Y_0$ abruptly shrinks to zero
when $\tilde w \geq \tilde w_\mathrm{max} \equiv 
\beta/\alpha^2$, or in other words for $w\geq w_\mathrm{max} \equiv
\lambda \eta^2 /b$: the range of variations for the state parameter
is automatically constrained, as in the ordinary Witten case
\cite{neutralPP92}.

The finite range of variation of the state parameter can
be understood in the following way. Imagine a region along the
string network where a statistical fluctuation on the phase gradient
implies the condensate should form with a very large value of $w$.
This gives the would-be condensate enough momentum to pass over
the potential barrier (\ref{calV}), and hence blocks the instability to
effectively take place until the fluctuation goes to a more reasonable
value below the maximum $(\partial\xi)^2 \leq w_\mathrm{max}$.

These equations are derivable from
the dimensionless action $\mathcal{S}_+$, where
\begin{widetext}
\begin{equation}
\mathcal{S}_\pm = \int \left\{ \dot X^2 + \dot Y^2 +\tilde w \dot P^2
+\frac{\dot Q^2}{\rho^2} \pm \tilde w\left[X^2P^2 +\left(\alpha+P\right)^2 Y^2
\right] +\frac{Q^2 X^2 + \left(Q-n\right)^2 Y^2}{\rho^2} + \frac12 \beta
\left(X^2+Y^2-1\right)^2\right\} \,\rho\,\dd\rho,
\label{Sadim}
\end{equation}
\end{widetext}
which is used to produce the numerical solutions shown on 
Fig.~\ref{fig:fields} that are discussed below. The quantities
$\mathcal{S}_\pm$ serve to define the energy per unit length
and tension through
\begin{equation}
U=2\pi \eta^2 \mathcal{S}_+ \ \ \ \ \hbox{and} \ \ \ \ \ 
T=2\pi \eta^2 \mathcal{S}_-.
\label{UTnum}
\end{equation}

We also derive the currents in terms of dimensionless variables as
\begin{equation}
I=\frac{\eta}{e} \pi\sqrt{2\tilde w} \int \left(\alpha
+ P\right) Y^2 \rho\,\dd\rho,
\label{Idimless}
\end{equation}
It is shown on Fig. \ref{fig:IthI} as functions of $\tilde w$.  The
limit provided by Eq. (\ref{Y0max}) compares with our numerical calculations
in the sense that the would-be current (\ref{Idimless})
obtained in Sec.~\ref{sec:eosA}
abruptly vanishes when $\tilde w$ exceeds the critical value above which
the condensate does not form at all. The other currents, i.e., the constraints
stemming from Eqs. (\ref{Ith}) and (\ref{Jfields}), are numerically verified
to hold, hence ensuring our field functions to solve their equations of motion.

Eqs.~(\ref{Sadim}) and (\ref{UTnum}) permit to show explicitely,
using the asymptotic behaviors derived above for $\sigma$, that
the energy and tension are both well behaved at the would-be phase
frequency threshold $w\to 0$. In terms of dimensionless variables,
we have, for $\rho\gg 1$, that  $Y(\rho)$ behaves as 
$Y\sim f(\tilde w) \tilde w^{n/2-1/4}
\ex^{-\alpha\sqrt{\tilde w}\rho}/\sqrt{\alpha\rho}$, where $f(\tilde w)$ is
an unknown function of $\tilde w$ whose behavior for small values of
the state parameter $\lim_{\tilde w\to 0} f(\tilde w)$ is a constant.

\begin{figure}[t]
\centering
\includegraphics[width=8cm,clip]{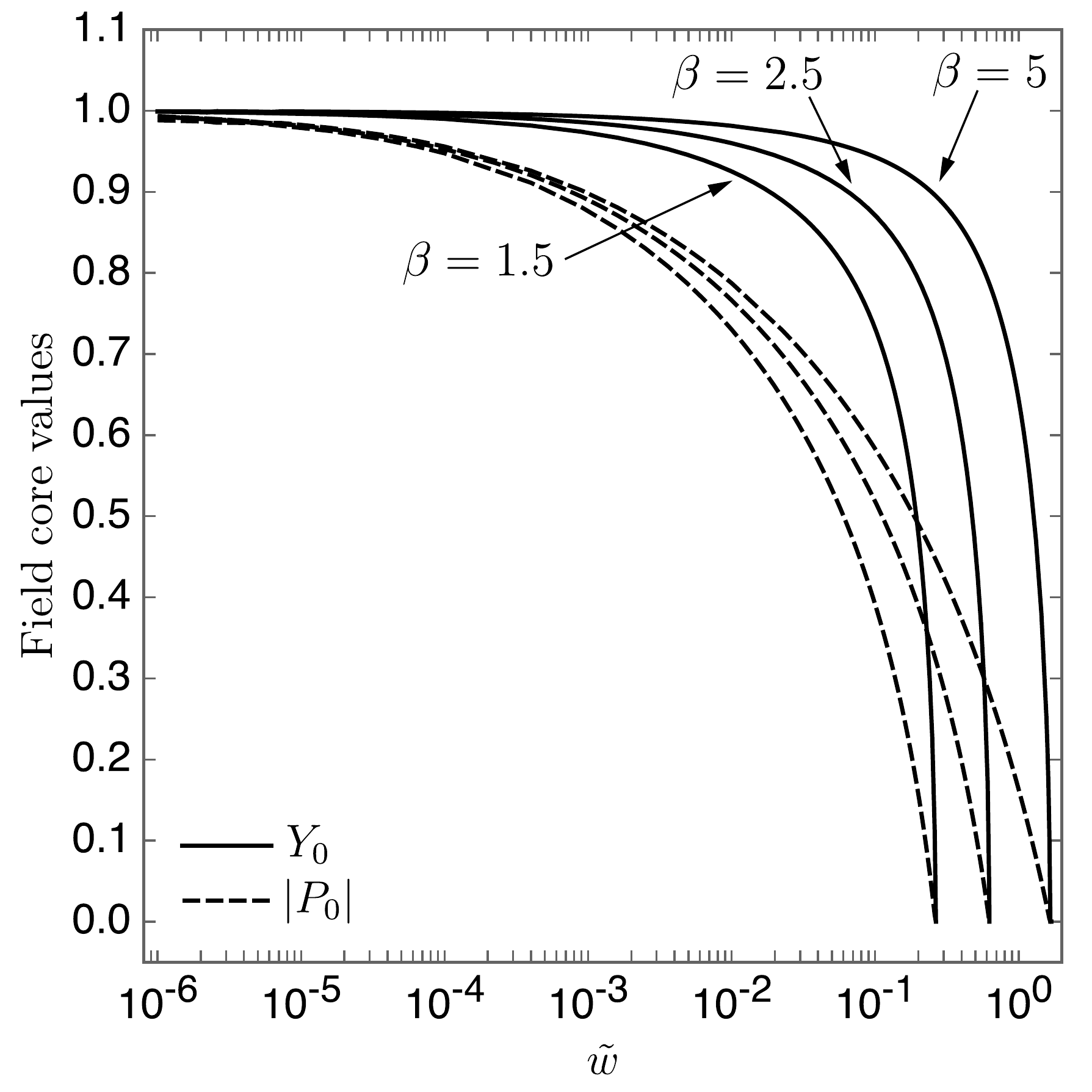}
\caption{Values of the condensate function $P_0(0)$ and $Y_0(0)$
in the string core ($\rho=0$) as functions of the rescaled state parameter
$\tilde w$ for $\alpha=1$ and various values of $\beta$ (same as on
Fig.~\ref{fig:UT}).}
\label{fig:Y0P0}
\end{figure}

Now, in this small $w$ regime, it is a simple matter to evaluate the
leading behavior of the integrated quantities, as most of the field
hardly depend on $w$: as shown on Fig.~\ref{fig:Y0P0}, the condensate
value at the string core and the current gauge function $P$, as well
as the background fields $X$ and $Q$, are essentially independent
of $w$. The only term that really matters for the variation of the
integrals with $w$ is the asymptotic behavior of the current carrier
$\sigma$: as in the ordinary Witten case, the condensate tends to
spread around the string around the phase frequency threshold,
\ie here in the almost chiral case. Thus, assuming the asymptotic
behavior to hold from a distance $\rho_\mathrm{M}$ on,
the dominant contribution
$\Delta$ comes from the $Y$ terms in Eq.~(\ref{Sadim}), namely
\begin{eqnarray}
\Delta_\pm &=& \int_{\rho_\mathrm{M}}^\infty 
\left[\dot Y^2 \pm \tilde w \left( \alpha+P\right)^2 Y^2
+\frac{\left(Q-n\right)^2}{\rho^2} Y^2 \right. \cr
&& \left. \hskip1cm +\beta\left(X^2-1\right) Y^2+
\frac12 \beta Y^4\right] \rho\,\dd\rho.
\nonumber
\end{eqnarray}
For $\rho > \rho_\mathrm{M}$, one can further make the
assumption that the other fields have reached their asymptotic
regime, namely we can set $(P,Q)\to 0$ and $X\to 1$, so the
only important contributions end up being
$$\Delta = \int_{\rho_\mathrm{M}}^\infty 
\left[\dot Y^2 \pm \tilde w \alpha^2 Y^2
+\frac{n^2}{\rho^2} Y^2+
\frac12 \beta Y^4\right] \rho\,\dd\rho,
$$
which can be explicitly calculated. Neglecting irrelevant
constant terms and keeping only the leading contributions,
this gives, for $n=1$ (the general case leads to similar
conclusions but is merely more involved and, as discussed
above, not relevant to the current discussion since the
type II vortices here considered are unstable for $n>1$,
splitting into $n$ unit winding vortices)
$$
\Delta \sim A \pm B \tilde w + C \sqrt{\tilde w}
+ D \tilde w \ln \tilde w,
$$
where $A$, $B$, $C$ and $D$ can be evaluated as asymptotic
integrals over the fields that do not depend on $\tilde w$.

\begin{figure}[t]
\centering
\includegraphics[width=8cm,clip]{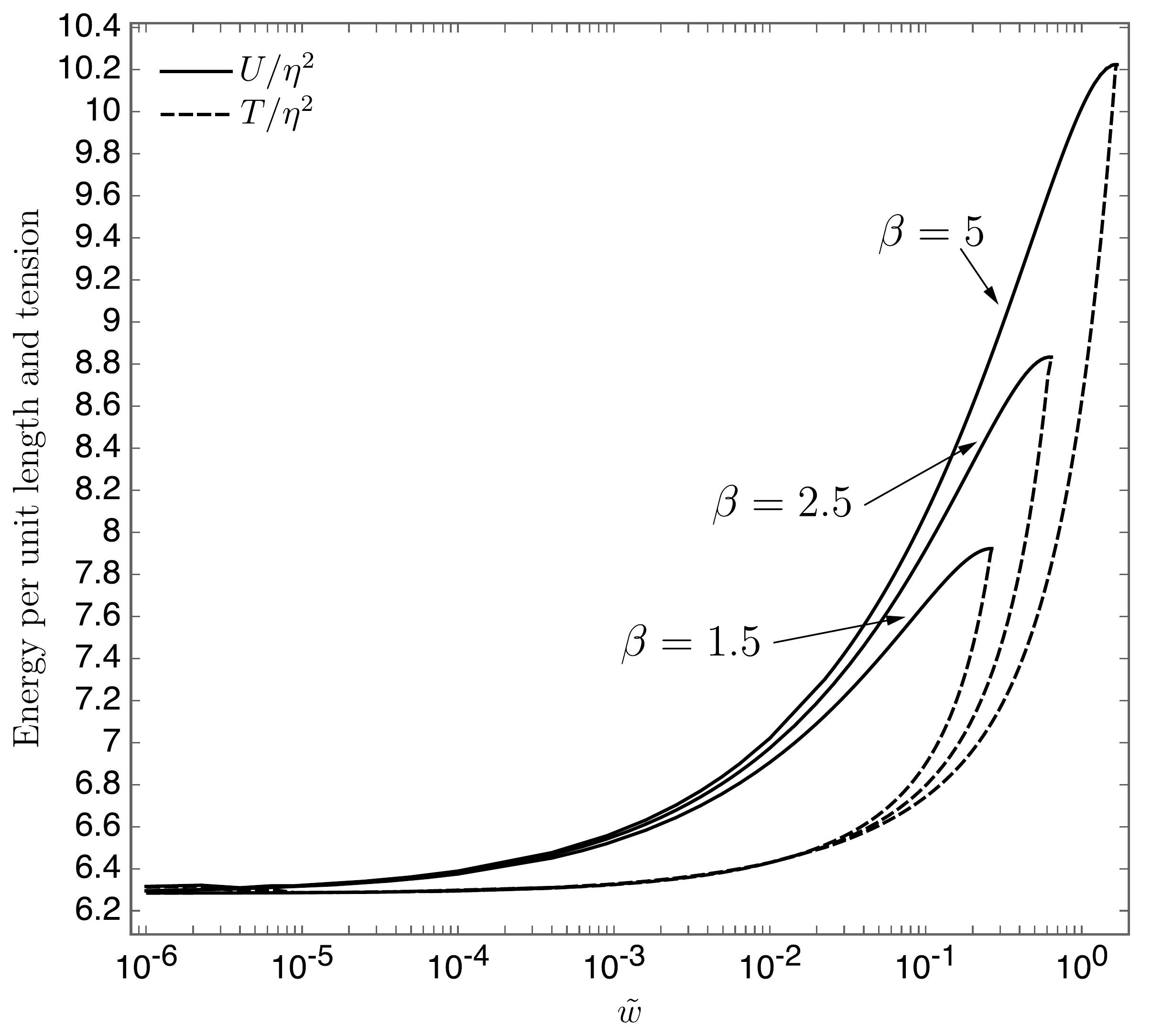}
\caption{Energy per unit length $U$ (full lines) and tension
$T$ (dashed lines) of the semi-local string in units of the
squared Higgs vev $\eta^2$
for $\alpha=1$ as on Fig. \ref{fig:fields}, and different values
of $\beta$ as indicated on the curves.
This shows explicitly the absence of a phase frequency threshold
at $\tilde w=0$, \ie the null current limit is perfectly regular. The
functions end abruptly for a maximum value of $\tilde w$, as
indicated in Eq.~(\ref{Y0max}) after which the condensate identically
vanished. It is also seen that $U$ and $T$ vary in the same way
with $w$ for all values of $w$, so that $\dd T/\dd U >0$, and hence
the longitudinal perturbation velocity $\cL^2$ is always negative,
signaling an unstable behavior of the string seen as a macroscopic
object; the relevant string evolution presumably leads to a chiral
behavior independently of the initial value of the cosmological $w$
distribution at the string network formation time.}
\label{fig:UT}
\end{figure}

What makes $U$ different from $T$ as functions of $\tilde w$ is,
in the above expression, the second term involving $B$. In the limit
$\tilde w\to 0$, this term rapidly becomes negligible, and the dominant
contribution thus implies that $U$ and $T$ evolve in similar ways
with respect to $\tilde w$, the unique parameter describing the
string state. As a result, variations of the tension with the energy
per unit length are always positive, so the longitudinal perturbation
velocity 
\begin{equation}
c^2_{_\mathrm{L}}\equiv - \frac{\dd T}{\dd U} \leq 0,
\label{cL}
\end{equation}
is negative in the limit $\tilde w\to 0$. Numerical calculation shown
in Fig.~\ref{fig:UT}
for the full range of available variations of $\tilde w$ shows that
in fact, Eq.~(\ref{cL}) is valid for all possible states attainable by
the strings under scrutiny here.

\section{Discussion and conclusion}
\label{conclude}

We have investigated a specific model of embedded type II
gauged vortices coming from the gauging of a U(1) subgroup
of an otherwise global SU(2). When the U(1) symmetry is
broken through a Higgs doublet acquiring a nonvanishing
vacuum expectation value, another component of the same
doublet can be excited because of a well-known condensate instability.
This leads to possible current-carrying string states as the
phase gradient of the carrier part of the doublet varies along the
string: at least at the time when the condensate forms, variations
from one point to another are subject to fluctuations over distances
larger than the correlation length, \ie the inverse mass of the
Higgs field.

Because type II vortices exhibit only a single unstable mode,
it was suggested in \cite{GaraudVolkov08} that those thus formed
could be stable provided they appear as sufficiently
small loops so that the unstable long wavelength microscopic
perturbations do not take over the dynamics. It remained to
understand whether these loops could be macroscopically stable,
and this requires that we solve the internal string structure in
order to be able to integrate over the irrelevant degrees of freedom.
This is achieved by means of a numerical integration of the
field equations and a calculation of the relevant integrated quantities
forming the stress-energy tensor, to be later coupled to gravity,
and the currents. We found that contrary to the original U(1)$\times$U(1)
Witten model \cite{Witten85,neutralPP92,enon0PP92} for
which a large region of stability with timelike, lightlike and
spacelike currents could be identified, here only a spacelike
current could be constructed. This relies on the fact that the
condensate is essentially a massless Goldstone mode, so that
any timelike excitation would be energetically favored to move
away from the string. The lightlike limit, however, appears to
be reasonably well-defined.

We obtained another crucial difference with the usual current-carrying
string models: the spacelike current configurations happen to be
unstable with respect to longitudinal (sound-wave like) perturbations.
As a result, our investigation closes the window of possible
stability zones opened in Ref.~\cite{GaraudVolkov08}, and we are led
to the definite conclusion that type II vortices cannot form, or if they
do, they will spontaneously decay in such a way that their cosmological
relevance is vanishing.

Let us finally point out that the non-current carrying semi-local strings
share some features with BPS D-term string solutions \cite{bdr}, in particular 
the latter possess a zero mode - very similar to semi-local strings. 
While the zero mode can also be excited \cite{as} one might wonder whether any of the results obtained
in our paper would also be valid in this case and whether this could
lead to any consequence
on inflationary models rooted in String Theory.

\acknowledgments

BH gratefully acknowledges support within the framework of the DFG Research
Training Group 1620 {\it Models of gravity}. PP would like to thank support from
Jacobs University (Bremen -- Germany). This research was supported in
part by Perimeter Institute for Theoretical Physics (Waterloo, ON --
Canada).

\end{document}